\documentclass[prd,aps,superscriptaddress,floatfix,
showkeys,showpacs,nofootinbib,twocolumn,10pt]{revtex4-2}
\pdfoutput=1
\usepackage{keyval,amsfonts,slashed,bm}
\usepackage{graphicx,textcomp}
\usepackage[colorlinks=true,linktocpage=true,linkcolor=blue,citecolor=blue]{hyperref}
\usepackage{epsfig,amsmath,amssymb}
\usepackage{subfig}
\usepackage{caption}
\usepackage{graphicx}

\usepackage{color}
\definecolor{blue(munsell)}{rgb}{0.2, 0.3, 0.69}
\definecolor{coquelicot}{rgb}{1.0, 0.22, 0.0}
\newcommand{\ba}{\begin{eqnarray}}
\newcommand{\ea}{\end{eqnarray}}

\newcommand{\nn}{\nonumber\\}

\definecolor{sinopia}{rgb}{0.8,0.25,0.04}

\definecolor{greenopia}{rgb}{0.3,0.65,0.14}

\makeatletter
\newsavebox{\@brx}
\newcommand{\llangle}[1][]{\savebox{\@brx}{\(\m@th{#1\langle}\)}%
	\mathopen{\copy\@brx\kern-0.5\wd\@brx\usebox{\@brx}}}
\newcommand{\rrangle}[1][]{\savebox{\@brx}{\(\m@th{#1\rangle}\)}%
	\mathclose{\copy\@brx\kern-0.5\wd\@brx\usebox{\@brx}}}
\makeatother

\begin{document}

\title{Heavy quark dynamics via Gribov-Zwanziger approach}
\author{Sumit}
\email{sumit@ph.iitr.ac.in}
\affiliation{Department of Physics,	Indian Institute of Technology Roorkee, Roorkee 247667, India}
	\author{Arghya Mukherjee}
\email{arbp.phy@gmail.com}
\affiliation{Ramakrishna Mission Residential College (Autonomous), Narendrapur, Kolkata-700103, India}
\author{Najmul Haque}
\email{nhaque@niser.ac.in}
\affiliation{School of Physical Sciences, National Institute of Science Education and Research,
	An OCC of Homi Bhabha National Institute, Jatni-752050, India}
\author{Binoy Krishna Patra}
\email{binoy@ph.iitr.ac.in}
\affiliation{Department of Physics,
	Indian Institute of Technology Roorkee, Roorkee 247667, India}

\begin{abstract}
In this work, we investigate the momentum-dependent drag and diffusion coefficient of heavy quarks $(\mathrm{HQs})$ moving in the quark-gluon plasma (QGP) background. The leading order scattering amplitudes required for this purpose have been obtained using the  Gribov-Zwanziger propagator for the mediator gluons to incorporate the non-perturbative effects relevant to the phenomenologically accessible temperature regime. The drag and diffusion coefficients so obtained have been implemented to estimate the temperature and momentum dependence of the energy loss of the $\mathrm{HQ}$ as well as the temperature dependence of the specific shear viscosity ($\eta/s$) of the background medium. Our results suggest a higher energy loss of the propagating $\mathrm{HQ}$ compared to the perturbative estimates, whereas the $\eta/s$ is observed to comply with the AdS/CFT estimation over a significantly wider temperature range compared to the perturbative expectation.    

		
\end{abstract}



\maketitle 
		
\section{Introduction}\label{Int}
The ultimate aim of the ongoing experiments, namely the Relativistic Heavy Ion Collider (RHIC) at Brookhaven National Laboratory (BNL) and Large Hadron Collider (LHC) at the European Council for Nuclear Research (CERN), is to create and study the new state of matter where bulk properties of this matter are governed by light quarks and gluons~\cite{Shuryak:2004cy, Jacak:2012dx}. It is now widely proven that this new state, which is the deconfined state of quarks and gluons known as strongly interacting quark-gluon plasma (sQGP), is created in these high energies nuclei collisions~\cite{Gyulassy:2004zy}. The models which successfully describe the space-time evolution of QGP fireball are governed by relativistic hydrodynamic models~\cite{Teaney:2000cw, Huovinen:2001cy, Nonaka:2006yn, Song:2007fn, Luzum:2008cw, Qiu:2011hf, Pang:2012uw, Gale:2012rq}, which gives information that the shear viscosity to entropy density ($\eta/s$) ratio of produced QGP is very small. Also, the experimental data analysis at RHIC suggests that $\eta/s \approx 0.1-0.2$~\cite{Bernhard:2019bmu, JETSCAPE:2020mzn} which is a strong indicator that the produced QGP in these collisions is strongly coupled because for a strongly coupled system $\eta/s$ is small. For a weakly coupled system, this ratio is large. One of the essential ways to characterize the properties of sQGP is by using hard probes, which are created in the initial stages of these highly energetic collisions, as their production requires a large momentum transfer. One of the promising hard probes is offered by heavy quarks $\mathrm{(HQs)}$, mainly charm and a bottom quark, because they do not constitute the bulk constituents of the matter and because of their large mass compared to the temperature scale generated in these ultrarelativistic heavy ion collisions (uRHICs)~\cite{Rapp:2009my}. $\mathrm{HQs}$ travel in the expanding medium as generated after these collisions and interact with the light particles of the medium. However, their number is most likely to be conserved because of their considerable $M/T$ ratio where $M$ is the mass of the HQ and $T$ is the temperature of the medium. Thus, $\mathrm{HQs}$ can experience the complete evolution of the QGP, and as they are produced in out-of-equilibrium, they are expected to retain their memory of interaction with plasma evolution~\cite{Rapp:2009my, Andronic:2015wma, Prino:2016cni, Aarts:2016hap, Greco:2017rro, Zhao:2020jqu}. Also, their thermal production and annihilation can be ignored. In a perturbative QCD (pQCD) framework, the thermalization time of heavy quark ($\mathrm{HQ}$) has been estimated which is of the scale of $10-15$ fm/c for charm quark and the scale of $25-30$ fm/c for bottom quark~\cite{Rapp:2009my, Moore:2004tg,vanHees:2005wb, Cao:2011et} for the temperature scales required for QGP formed in RHIC and LHC experiments. Nevertheless, since the lifetime of QGP is around $4-5$ fm/c at RHIC and $10-12$ fm/c at LHC. Therefore, one should not expect the complete thermalization of $\mathrm{HQs}$ in uRHICs. For the small momentum exchange, the multiple scattering of $\mathrm{HQ}$ in a thermalized medium can be dealt with as a Brownian motion, and Boltzmann equation in that approximation reduced to Fokker-Planck equation~\cite{Rapp:2009my, Moore:2004tg, Svetitsky:1987gq, GolamMustafa:1997id} which constitutes a simplified version of in-medium dynamics. This method has been widely used~\cite{Moore:2004tg,vanHees:2005wb, Cao:2011et, GolamMustafa:1997id,vanHees:2007me, Cao:2012au, Young:2011ug, Alberico:2011zy, Akamatsu:2008ge, He:2012df} to study the experimental observables such as nuclear modification factor ($R_{AA}$)~\cite{PHENIX:2006iih, STAR:2006btx, PHENIX:2005nhb, ALICE:2012ab} and elliptic flow ($v_{2}$)~\cite{PHENIX:2006iih} for nonphotonic electron spectra.   

$\mathrm{HQ}$ production has been explored in the perturbative QCD approach up to the next-to-leading order (NLO). In the perturbative realm, before the first experimental result, it was anticipated that their interaction with the medium particles could be described using a perturbative QCD (pQCD) technique, which leads to the expectation of small suppression of the final spectra and a small value of the elliptic flow. Nevertheless, experimental results come with a surprise in which the spectrum of nonphotonic electrons coming from the heavy quark decays has been observed in Au-Au collision at $\sqrt{s}=200$GeV at RHIC~\cite{PHENIX:2006iih, STAR:2006btx, PHENIX:2005nhb}. This result shows a relatively small $R_{AA}$ and a large value of elliptic flow $v_{2}$, which clearly indicates that there is a strong correlation between the $\mathrm{HQ}$ and medium constituents, which is beyond the pQCD explanations~\cite{Moore:2004tg,vanHees:2005wb, vanHees:2004gq}. This motivates one to go beyond pQCD to tackle the problem in a non-perturbative manner. One of the approaches is to consider the non-perturbative contribution~\cite{vanHees:2007me} from the quasi-hadronic bound states with subsequent hadronization from coalescence and fragmentation~\cite{Greco:2003vf, Greco:2003xt}. Another method consists of the hard thermal loop (HTL) in the pQCD framework to calculate the debye mass and running coupling~\cite{Gossiaux:2008jv, Alberico:2011zy}. This technique includes the non-perturbative contributions through the inclusion of thermal mass $\sim g(T) T$ where the running coupling has been fitted via lattice thermodynamics~\cite{Plumari:2011mk, Berrehrah:2014kba}. All these models are built upon the assumption that collisional energy loss serves as the predominant process in the low-momentum range of charm spectra~\cite{Cao:2011et, Cao:2012au, Cao:2016gvr}, $p_{T}\lesssim (3-5)$ $m_{\mathrm{HQ}}$. On the other hand, at high $p_{T}$, the radiative effects are dominating even collisional ones can not be disregarded although~\cite{Djordjevic:2011dd, Gossiaux:2010yx, Das:2010tj}. In the low transverse momentum region, the collisional energy loss process dominated because of the effect that the phase space for the in-medium induced gluon radiation is constrained because of $\mathrm{HQ}$ mass, i.e., ``dead-cone effect''~\cite{Dokshitzer:2001zm, Abir:2012pu}. However, now, at LHC experiments, heavy meson spectra can be observed around $30$ GeV. At such high $p_{T}$, even $\mathrm{HQs}$ become ultrarelativistic, and thus, radiative energy loss effects become important.

Energetic particles traversing the QCD medium suffer energy loss through the elastic process and gluon-bremsstrahlung. The drag and diffusion of $\mathrm{HQs}$ cause them to lose their energies in the medium. Much work has been done in the literature to study the energy loss of $\mathrm{HQs}$. $\mathrm{HQ}$ energy loss due to hard and soft collision processes has been studied in~\cite{Braaten:1991jj, Braaten:1991we} and for radiative processes in~\cite{Mustafa:1997pm, Mustafa:2004dr, Qin:2007rn, Abir:2012pu}. Recently, the soft contribution of the parton energy loss has been studied within a chiral imbalance in~\cite{Ghosh:2023ghi}. Many studies have been done recently in literature to understand the $\mathrm{HQ}$ dynamics like $\mathrm{HQ}$ potential~\cite{Thakur:2016cki, Thakur:2020ifi, Sebastian:2022sga}, spectral properties~\cite{Thakur:2021vbo}, transport coefficients without~\cite{Das:2012ck, Das:2010tj, Hong:2023cwl} and with bulk viscous medium~\cite{Shaikh:2021lka, Shaikh:2023qei}. Transport phenomenon has been studied for various other cases like Polyakov loop plasma~\cite{Singh:2018wps}, semi-QGP~\cite{Singh:2019cwi}, and memory effects in $\mathrm{HQ}$ dynamics~\cite{Ruggieri:2022kxv, Khowal:2021zoo}.   

One of the other approaches to studying the non-perturbative phenomenon in $\mathrm{HQ}$ dynamics can be made by using the Gribov-Zwanziger~\cite{Gribov:1977wm, Zwanziger:1989mf} technique. This method improves the infrared dynamics of QCD through a scale of the order of $g^{2} T $, which is known as the magnetic scale of the theory. This model deals with the non-perturbative resummation of the theory, having a mass parameter that captures the non-perturbative essence of the theory. For some good reviews, one can look at~\cite{Dokshitzer:2004ie, Vandersickel:2012tz}. This approach has been extended by including the impact of a local composite operator, which consists of a mass term of the order of electric scale $g(T) T$. For more details on this extended Gribov-Zwanziger method, see some of the recent works in~\cite{Dudal:2008sp, Capri:2014bsa, Capri:2015ixa, Capri:2015nzw, Capri:2016aqq, Dudal:2019ing, Dudal:2017kxb, Gotsman:2020ryd, Gotsman:2020mpg, Justo:2022vwa, Gracey:2010cg} and references therein. This extended approach with mass term inclusion in the propagator gives results that are very promising with lattice calculations in the infrared domain, as shown in~\cite{Dudal:2008sp}. Also, at zero temperature, it has been shown in~\cite{Tissier:2010ts, Mintz:2018hhx} that this mass term inclusion in gluon and ghost propagator of the usual Faddeev-Popov quantization is in excellent agreement with lattice results. More details about this can be found in a recent review~\cite{Pelaez:2021tpq}.  

Without any mass term, this scheme has been quite successful in describing the QCD thermodynamics when a comparison with lattice simulations has been made in~\cite{Fukushima:2013xsa}. Also, the other exciting studies which have been done in the recent literature explore quark dispersion relations~\cite{Su:2014rma}, connection between Gribov quantization and confinement-deconfinement transition~\cite{Kharzeev:2015xsa}, the transport coefficients~\cite{Florkowski:2015dmm, Florkowski:2015rua, Jaiswal:2020qmj}, the dilepton production rate has been calculated along with quark number susceptability~\cite{Bandyopadhyay:2015wua}, screening masses of mesons~\cite{Sumit:2023hjj} and electromagnetic debye mass~\cite{Bandyopadhyay:2023yjp}, which gives some interesting results of the said observables. In the context of $\mathrm{HQ}$ phenomenology, which we are interested in here, the heavy quarkonium potential has been calculated using this method in~\cite{Bandyopadhyay:2023yjp, Wu:2022nbv, Debnath:2023dhs}, the collisional energy loss of $\mathrm{HQs}$ has been estimated in~\cite{Debnath:2023zet} incorporating the formalism of Wong equations and $\mathrm{HQ}$ diffusion coefficient using Langevin dynamics has been studied in~\cite{Madni:2022bea}. 

In this work, we explored the finite momentum-dependent~\cite{Mazumder:2011nj} drag and diffusion coefficient of $\mathrm{HQs}$ using the Gribov gluon propagator. Earlier in the literature, carrying forward the calculation of drag and diffusion coefficient using the perturbative approaches, there was a need to set some infrared scale to tackle the infrared divergences that arise mainly in $t-$channel exchange diagrams. The main advantage of this method is that one does not need any infrared cut-off to put by hand in the matrix element calculation; instead, it comes automatically in the model calculations. Also, as discussed earlier, the ratio of shear viscosity to entropy density ratio, which is an essential observable to quantify the nature of QGP, has been studied earlier using the perturbative methods~\cite{Mazumder:2013oaa}. It was found that the inclusion of radiative effects in the calculation improves this ratio significantly. 

The work is organized as follows: Following this concise introduction in Section~\ref{Int}, we will delve into the conventional formalism for calculating the drag and diffusion coefficients of $\mathrm{HQs}$. This will be accomplished using the widely recognized Fokker-Planck method, as detailed in Section~\ref{sec._2}. In this section, we will discuss the scattering of $2\rightarrow 2$ collisional process as well as the $2\rightarrow3$ radiative process. We give the required matrix element calculation, which has been done using the Gribov propagator. Section~\ref{sec.3} focuses on our results for the drag and diffusion coefficient as estimated using the Gribov propagator. A critical observable $\eta/s$ which is required to understand the nature of the QGP, i.e., whether a medium behaves like a weakly coupled or strongly coupled system, has been plotted using the Gribov propagator. Also, the energy loss of $\mathrm{HQs}$ has been discussed within this model for the charm and bottom quarks while traversing the medium. In section ~\ref{sec.4}, we summarize the paper and give further directions.


\section{Formalism: Drag and diffusion coefficients}\label{sec._2}
As discussed earlier, the motion of $\mathrm{HQs}$ in the QCD medium can be considered as a Brownian motion and is well described by the Fokker-Planck equation~\cite{Svetitsky:1987gq, GolamMustafa:1997id}
\begin{equation}\label{FP_equa.}
\frac{\partial f_{\mathrm{HQ}}}{\partial t}=\frac{\partial}{\partial p_i}\left[A_i(\bm{p}) f_{\mathrm{HQ}}+\frac{\partial}{\partial p_j}\left[B_{i j}(\bm{p}) f_{\mathrm{HQ}}\right]\right] \, ,
\end{equation}
where $f_{\mathrm{HQ}}$ represents the $\mathrm{HQ}$ momentum distribution in the medium. In this approach, the interaction of $\mathrm{HQ}$ with the medium constituent particles, which are light quarks, anti-quarks, and gluons, is encoded in the drag and diffusion tensors $A_{i}$ and  $B_{ij}$ respectively, which naturally arise from the momentum expansion of the collision integral of the Boltzmann transport equation (BTE)~\cite{Svetitsky:1987gq}. In the following, we briefly discuss the essential steps to obtain the drag and diffusion tensor of the $\mathrm{HQ}$. For clarity, the collisional and the radiative contributions are discussed in separate subsections.      

\subsection{Collisional Processes:}
Let us start with the two-body elastic scattering process:
$\mathrm{HQ}(P)+l(Q) \rightarrow$ $\mathrm{HQ}\left(P^{\prime}\right)+l\left(Q^{\prime}\right)$,  where $l$ denotes light particles viz. light quarks, anti-quarks, and gluons. Here, the four-momentum of the $\mathrm{HQ}$ and the constituent particle before the collision is represented by $P = (E_{p},\bm{p})$ and $Q = (E_{q},\bm{q})$  respectively. The corresponding four-momentum after the collision is denoted with primes.
Note that, in the case of the $\mathrm{HQ}$, the energy is given by $E_{p} = (|\bm{p}|^{2}+m_{\mathrm{HQ}}^{2})^{1/2}$ whereas the light particles are considered to be massless with $E_{q} = |\bm{q}|$. The  drag and the diffusion tensor that govern the dynamics  of the $\mathrm{HQ}$ in the QGP medium can be related to the $2 \rightarrow 2$ scattering  amplitude as~\cite{Svetitsky:1987gq}
\begin{eqnarray}
A_i&= & \frac{1}{2 E_p} \int \frac{d^3 \bm{q}}{(2 \pi)^3 2 E_q} \int \frac{d^3 \bm{q}^{\prime}}{(2 \pi)^3 2 E_{q^{\prime}}} \int \frac{d^3 \bm{p}^{\prime}}{(2 \pi)^3 2 E_{p^{\prime}}} \frac{1}{g_{\mathrm{HQ}}} \nonumber\\
& \times& \sum\left|\mathfrak{M}_{2 \rightarrow 2}\right|^2(2 \pi)^4 \delta^4\left(P+Q-P^{\prime}-Q^{\prime}\right) f_k({E_{q}}) \nonumber\\
&\times &\left[1+a_k f_k\left({E_{q^{\prime}}}\right)\right]\left[\left(\bm{p}-\bm{p}^{\prime}\right)_i\right]=\llangle\left(\bm{p}-\bm{p}^{\prime}\right)_i\rrangle , \hspace*{-.0cm}  
\end{eqnarray}
\begin{equation}
\begin{aligned}
B_{i j}= & \frac{1}{2}\llangle[\Big]\left(\bm{p}-\bm{p}^{\prime}\right)_i\left(\bm{p}-\bm{p}^{\prime}\right)_j
\rrangle[\Big] \, .
\end{aligned}\label{diffusion_definition}
\end{equation} 
The expressions above indicate that the drag force represents the thermal average of the momentum transfer $\left(\bm{p}-\bm{p}^{\prime}\right)$ resulting from interactions. On the other hand, momentum diffusion quantifies the average square of the momentum transfer. In these expressions, $g_{\mathrm{HQ}}$ represents the statistical degeneracy factor of the $\mathrm{HQ}$, and the subscript $k$ denotes the particle species in the medium. The quantity $a_{k} = 1,-1$ represents respectively the near-equilibrium Bose-Einstein and the Fermi-Dirac distributions denoted in general as $f_{k}$. The delta function enforces the energy-momentum conservation. The computation of the matrix amplitude $\mathfrak{M}_{2 \rightarrow 2}$  for the allowed $ 2 \rightarrow 2$ scattering processes will be discussed in the following subsection. It should be noted that the drag force depends only on $\mathrm{HQ}$ momentum. Thus, one can decompose it as
\begin{equation}
A_{i}=p_i A (p^2) \, , \quad \quad  A=\llangle 1\rrangle-\frac{\llangle\bm{p} \cdot \bm{p}^{\prime}\rrangle}{p^2} \,.
\end{equation}
where $p^{2} = |\bm{p}|^{2}$ and $A$ is the drag coefficient of $\mathrm{HQ}$. Similarly, one can decompose the diffusion tensor $B_{i j}$ in terms of transverse and longitudinal components with respect to $\mathrm{HQ}$ momentum as
\begin{equation}
B_{i j}=\left(\delta_{i j}-\frac{p_i p_j}{p^2}\right) B_0\left(p^2\right)+\frac{p_i p_j}{p^2} B_1\left(p^2\right)\, ,
\end{equation}
where the transverse diffusion coefficient $B_0$ and longitudinal diffusion coefficient $B_1$ take the following forms
\begin{equation}
\begin{aligned}
& B_0=\frac{1}{4}\left[\llangle[\big] p^{\prime 2}\rrangle[\big]-\frac{\llangle[\big]\left(\bm{p}^{\prime} \cdot \bm{p}\right)^2\rrangle[\big]}{p^2}\right]\, , \\
\end{aligned}
\end{equation}
\begin{equation}
\begin{aligned}
& B_1=\frac{1}{2}\left[\frac{\llangle[\big]\left(\bm{p}^{\prime} \cdot \bm{p}\right)^2\rrangle[\big]}{p^2} - 2\llangle[\big]\bm{p}^{\prime} \cdot \bm{p}\rrangle[\big]+p^2\llangle[\big] 1\rrangle[\big]\right] \,.
\end{aligned}
\end{equation}
One can study the kinematics of the $2 \rightarrow 2$ process in the center-of-momentum $\mathrm{(COM)}$ frame for simplification. The average of a generic function $F(\bm{p})$ in the $\mathrm{COM}$ frame can be written as~\cite{Svetitsky:1987gq, Kumar:2021goi}
\ba
\llangle F(\bm{p})\rrangle &= & \frac{1}{\left(512 \pi^4\right) E_p g_{\mathrm{HQ}}} \int_0^{\infty} q\, d q\left(\frac{s-m_{\mathrm{HQ}}^2}{s}\right) f_k\left(E_q\right) \nn
&\times&  \int_0^\pi d \chi \sin \chi \int_0^\pi d \theta_{\mathrm{cm}} \sin \theta_{\mathrm{cm}} \sum\left|\mathfrak{M}_{2 \rightarrow 2}\right|^2 \nn
&\times& \int_0^{2 \pi} d \phi_\mathrm{cm}\left[1+a_k f_k\left(E_{q^{\prime}}\right)\right] F(\bm{p}) \, .
\ea
where $\chi$ quantifies the angle between the incident $\mathrm{HQ}$ and the medium constituent particles in the laboratory frame, while $\theta_{\mathrm{cm}}$ and $\phi_{\mathrm{cm}}$ are respectively the zenith and azimuthal angles in the $\mathrm{COM}$ frame. The Mandelstam variables $s, t$ and $u$ are defined as follows
\ba
s & =& (P+Q)^{2} = \left(E_p+E_q\right)^2-\left(p^2+q^2+2 p q \cos \chi\right) \, , \nn
t & =& (P^{\prime}-P)^{2} = 2 p_{\mathrm{cm}}^2\left(\cos \theta_{\mathrm{cm}}-1\right) \, , \nn
u & =& (P^{\prime}-Q)^{2} = 2 m_{\mathrm{HQ}}^2-s-t \, .
\ea
Here $p_{\mathrm{cm}} = |\bm{p}_{\mathrm{cm}}|$ is the magnitude of the initial momentum of the $\mathrm{HQ}$ in the $\mathrm{COM}$ frame. The other quantity required in order to obtain the drag and diffusion coefficients is $\left(\bm{p}\cdot\bm{p}^{\prime}\right)$. In order to find this quantity, we need the Lorentz transformation that relates the laboratory frame and the $\mathrm{COM}$ frame via the relation $ \bm{p}^{\prime}=\gamma_{\mathrm{cm}}\left(\hat{\bm{p}}_{\mathrm{cm}}^{\prime}+\bm{v}_{\mathrm{cm}} \hat{E}_{\mathrm{cm}}^{\prime}\right),$ where  $\gamma_{\mathrm{cm}}=(E_p+E_q)/\sqrt{s}$ and the velocity in the $\mathrm{COM}$ is given by $\bm{v}_{\mathrm{cm}}=(\bm{p}+\bm{q})/(E_p+E_q)$. Now, the energy conservation dictates $\hat{p}_{\mathrm{cm}}^{\prime 2}=\hat{p}_{\mathrm{cm}}^2$. In the $\mathrm{COM}$ frame, $\hat{\bm{p}}_{\mathrm{cm}}^{\prime}$ can be decomposed as
$ \hat{\bm{p}}_{\mathrm{cm}}^{\prime}=  \hat{p}_{\mathrm{cm}}\left(\cos \theta_{\mathrm{cm}} \hat{\bm{x}}_{\mathrm{cm}}+\sin \theta_{\mathrm{cm}} \sin \phi_{\mathrm{cm}} \hat{\bm{y}}_{\mathrm{cm}}+\sin \theta_{\mathrm{cm}} \cos \phi_{\mathrm{cm}} \hat{\bm{z}}_{\mathrm{cm}}\right),$
where $\hat{p}_{\mathrm{cm}}=(s-m_{\mathrm{HQ}}^2)/(2 \sqrt{s})$ is the momentum and $\hat{E}_{\mathrm{cm}}=$ $\left(\hat{p}_{\mathrm{cm}}^2+m_{\mathrm{HQ}}^2\right)^{1 / 2}$ is the energy of the $\mathrm{HQ}$ in the $\mathrm{COM}$ frame. The axes $\hat{\bm{x}}_{\mathrm{cm}}, \hat{\bm{y}}_{\mathrm{cm}}$, and $\hat{\bm{z}}_{\mathrm{cm}}$ are defined in~\cite{Svetitsky:1987gq}. Utilizing the above definitions, one can obtain
\begin{equation}
\begin{aligned}
\bm{p} \cdot \bm{p}^{\prime} & = E_p E_p^{\prime}-\hat{E}_{\mathrm{cm}}^2+\hat{p}_{\mathrm{cm}}^2 \cos \theta_{\mathrm{cm}} \, .
\end{aligned}
\end{equation}

\subsection{Matrix Elements for $2 \rightarrow 2$ processes}
The leading order Feynman diagrams for $2 \rightarrow 2$ processes are shown in Fig.~\ref{fig:1}. There are three topologically distinct diagrams contributing to quark-gluon scattering shown in Figs.~\ref{fig1(a)}-\ref{fig1(c)} and one diagram for quark-quark or quark-anti-quark scattering shown in Fig.~\ref{fig1(d)}~\cite{Matsui:1985eu}. Note that each of the diagrams shown in Fig.~\ref{fig1(a)} and Fig.~\ref{fig1(d)} possesses a gluon propagator, which in the present work has been replaced with the Gribov-modified gluon propagator. The modified gluon propagator in the Landau gauge is given  as~\cite{Fukushima:2013xsa}
\begin{equation}\label{Gribov_prop}
D_{\mu\nu}^{ab}(P)=\delta^{a b} \left(\delta_{\mu \nu}-\frac{P_\mu P_\nu}{P^2}\right)\frac{P^2}{P^4+\gamma_{\mathrm{G}}^4}\,.
\end{equation}
where $\gamma_{\mathrm{G}}$ is the Gribov mass parameter which is generally derived from the one-loop or two-loop gap equation~\cite{Gracey:2010cg}. The matrix elements for the diagrams shown in Fig.~\ref{fig:1}, using the Gribov propagator, are given by
\begin{widetext}
\ba
 \mathfrak{M}_{(\mathrm{a})} &=& -g^2 \varepsilon_\mu(2) \varepsilon_\nu^{*}(4) f_{a b c}\left[g^{\mu \nu}\left(-Q-Q^{\prime}\right)^\rho+g^{\nu \rho}\left(2Q^{\prime}-Q\right)^\mu+g^{\rho \mu}\left(2 Q-Q^{\prime}\right)^\nu \right] \frac{(P^{\prime}-P)^{2}}{(P^{\prime}-P)^{4}+\gamma_{\mathrm{G}}^{4}} \bar{u}^{i}(3) \gamma_\rho \lambda_c u^{i}(1) \, , \nn
 \mathfrak{M}_{(\mathrm{b})} &=& -i g^2 \varepsilon_\mu(2) \varepsilon_\nu^{*}(4) \bar{u}^{i}(3) \gamma^\mu \lambda_a \frac{\slashed{P}+\slashed{Q}+m_\mathrm{HQ}}{\left(P+Q\right)^2-m_\mathrm{HQ}^2} \gamma^\nu \lambda_b u^{i}(1) \, , \nn
 \mathfrak{M}_{(\mathrm{c})} &=& -i g^2 \varepsilon_\mu(2) \varepsilon_\nu^{*}(4) \bar{u}^{i}(3) \gamma^\nu \lambda_b \frac{\slashed{P}^{\prime}-\slashed{Q}+m_\mathrm{HQ}}{\left(P^{\prime}-Q\right)^2-m_\mathrm{HQ}^2} \gamma^\mu \lambda_a u^{i}(1) \, , \nn
 \mathfrak{M}_{(\mathrm{d})} &=& i g^2 \bar{u}^{i}(3) \gamma^\mu \lambda_a u^{i}(1) \frac{(P^{\prime}-P)^{2}}{(P^{\prime}-P)^{4}+\gamma_{\mathrm{G}}^{4}} \bar{u}^{i}(4) \gamma_\mu \lambda_a u^{i}(2) \, .
\ea
Here, the abbreviated notations used are $\varepsilon_{\mu}(1) = \varepsilon_{\mu}(P,\zeta_{P}),\, \varepsilon_{\mu}(2) = \varepsilon_{\mu}(Q,\zeta_{Q}),\, \varepsilon_{\mu}(3) = \varepsilon_{\mu}(P^{\prime},\zeta_{P^{\prime}})$ and $\varepsilon_{\mu}(4) = \varepsilon_{\mu}(Q,\zeta_{Q^{\prime}})$ for gluon polarization vectors, $i$ denotes different flavors and $u(1) = u(P,s_{P})$ for quark spinors. The symbols $\lambda_{a}$ represent $SU(3)$ matrices normalized by $\text{Tr}(\lambda_{a}\lambda_{b}) = \frac{1}{2}\delta_{ab}$, satisfying $[\lambda_{a},\lambda_{b}] = i f_{abc} \lambda_{c}$ and $f_{abc}$ are the structure constants. The summation of squared matrix elements over the initial and final quark spin states transforms the quark spinors into projection operators as per the relation
\ba 
\sum_{s=1,2} u_\alpha^i(P, s) \bar{u}_\beta^i(P, s)&=& \left(\slashed{P}+m_{\mathrm{HQ}}^{i}\right)_{\alpha \beta} \, . 
\ea
During summation over gluon polarizations $\zeta_{r}$ where $r=1,2,3,4$, in order to avoid the contributions from the unphysical states, one can remove the terms containing  $\varepsilon_{\mu}(P,\zeta)P^{\mu}$. Thus the amplitude $\mathfrak{M}_{(\mathrm{a})}$ becomes
\ba
 \mathfrak{M}_{(\mathrm{a})} &=& -g^2 \varepsilon_\mu(2) \varepsilon_\nu^{*}(4) f_{a b c}\left[g^{\mu \nu}\left(-Q-Q^{\prime}\right)^\rho+g^{\nu \rho}\left(2Q^{\prime}\right)^\mu+g^{\rho \mu}\left(2 Q\right)^\nu\right] \frac{(P^{\prime}-P)^{2}}{(P^{\prime}-P)^{4}+\gamma_{\mathrm{G}}^{4}} \bar{u}^{i}(3) \gamma_\rho \lambda_c u^{i}(1) \, , \nn
\ea
\begin{figure}
\centering
\subfloat[\label{fig1(a)}]{\includegraphics[scale=1.2]{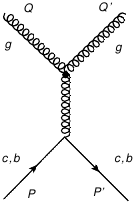}}
\quad \quad \quad
\subfloat[\label{fig1(b)}]{\includegraphics[scale=1.2]{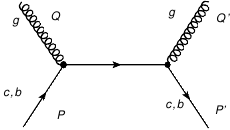}}
\quad \quad \quad
\subfloat[\label{fig1(c)}]{\includegraphics[scale=1.2]{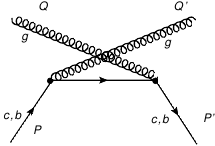}}
\quad \quad \quad
\subfloat[\label{fig1(d)}]{\includegraphics[scale=1.2]{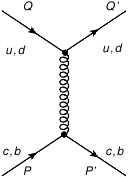}}
\caption{Feynman diagrams for $\mathrm{HQ}$ $2$ $\rightarrow$ $2$ processes with (a) gluon ($t-$channel), (b) gluon ($s-$channel), (c) gluon ($u-$channel), (d) light quark/anti-quark ($t-$channel).}%
\label{fig:1}%
\end{figure}
Now, one can do the trace over the Lorentz indices utilizing the relation
\begin{equation}
\begin{aligned}
\sum_{\zeta = 1,2} \varepsilon_{\mu}^{*}(P,\zeta)\varepsilon_{\nu}(P,\zeta) = - g_{\mu\nu} \, .
\end{aligned}
\end{equation}
The squared matrix elements can be conveniently written in terms of Mandelstam variables, 
which satisfy the relation $s+t+u=2m_{\mathrm{HQ}}^{2}$. After doing the summation over the spins, polarizations, and color indices, one would get the final expressions as follows
(i) for the process $\mathrm{HQ}(P)+g(Q) \rightarrow$ $\mathrm{HQ}\left(P^{\prime}\right)+g\left(Q^{\prime}\right) $, one obtains 
\ba
\left|\mathfrak{M}_{(a)}\right|^2 &=& g_{\mathrm{HQ}} g_g\left[32 \pi^2 \alpha^2 \frac{\left(s-m_{\mathrm{HQ}}^2\right)\left(m_{\mathrm{HQ}}^2-u\right)t^{2}}{\left(t^{2}+\gamma_{\mathrm{G}}^4\right)^2}\right] \, , \nn
\left|\mathfrak{M}_{(b)}\right|^2 &=& g_{\mathrm{HQ}} g_g\left[\frac{64 \pi^2 \alpha^2}{9} \frac{\left(s-m_{\mathrm{HQ}}^2\right)\left(m_{\mathrm{HQ}}^2-u\right)+2 m_{\mathrm{HQ}}^2\left(s+m_{\mathrm{HQ}}^2\right)}{\left(s-m_{\mathrm{HQ}}^2\right)^2}\right] \, , \nn
\left|\mathfrak{M}_{(c)}\right|^2 &=& g_{\mathrm{HQ}} g_g\left[\frac{64 \pi^2 \alpha^2}{9} \frac{\left(s-m_{\mathrm{HQ}}^2\right)\left(m_{\mathrm{HQ}}^2-u\right)+2 m_{\mathrm{HQ}}^2\left(m_{\mathrm{HQ}}^2+u\right)}{\left(m_{\mathrm{HQ}}^2-u\right)^2}\right] \, , \nn
\mathfrak{M}_{(a)} \mathfrak{M}_{(b)}^* &=& \mathfrak{M}_{(b)}^* \mathfrak{M}_{(a)} = g_{\mathrm{HQ}}  g_g\left[8 \pi^2 \alpha^2 \frac{\left(s-m_{\mathrm{HQ}}^2\right)\left(m_{\mathrm{HQ}}^2-u\right)+m_{\mathrm{HQ}}^2(s-u)}{\left(\frac{t^{2}+\gamma_{\mathrm{G}}^4}{t} \right)\left(s-m_{\mathrm{HQ}}^2\right)}\right] \, , \nn
\mathfrak{M}_{(a)} \mathfrak{M}_{(c)}^* &=& \mathfrak{M}_{(c)}^* \mathfrak{M}_{(a)}= g_{\mathrm{HQ}} g_g\left[8 \pi^2 \alpha^2 \frac{\left(s-m_{\mathrm{HQ}}^2\right)\left(m_{\mathrm{HQ}}^2-u\right)-m_{\mathrm{HQ}}^2(s-u)}{\left(\frac{t^{2}+\gamma_{\mathrm{G}}^4}{t} \right)\left(m_{\mathrm{HQ}}^2-u\right)}\right] \, , \nn
\mathfrak{M}_{(b)} \mathfrak{M}_{(c)}^* &=& \mathfrak{M}_{(b)}^* \mathfrak{M}_{(c)}= g_{\mathrm{HQ}} g_g\left[\frac{8 \pi^2 \alpha^2}{9} \frac{m_{\mathrm{HQ}}^2\left(4 m_{\mathrm{HQ}}^2-t\right)}{\left(s-m_{\mathrm{HQ}}^2\right)\left(m_{\mathrm{HQ}}^2-u\right)}\right] \, , \nn
\left|\mathfrak{M}_{(i)}\right|^2 &=& \left|\mathfrak{M}_{(a)}\right|^2+\left|\mathfrak{M}_{(b)}\right|^2+\left|\mathfrak{M}_{(c)}\right|^2+2 \mathcal{R} e\left\{\mathfrak{M}_{(a)} \mathfrak{M}_{(b)}^*\right\}+2 \mathcal{R} e\left\{\mathfrak{M}_{(b)} \mathfrak{M}_{(c)}^*\right\} +2 \mathcal{R} e\left\{\mathfrak{M}_{(a)} \mathfrak{M}_{(c)}^*\right\} \, ,
\ea
and (ii) for the process $\mathrm{HQ}(P)+lq(Q)/l\bar{q}(Q) \rightarrow$ $\mathrm{HQ}\left(P^{\prime}\right)+lq\left(Q^{\prime}\right)/l\bar{q}\left(Q^{\prime}\right)$, one obtains
\begin{equation}
\begin{aligned}
\left|\mathfrak{M}_{(d)}\right|^2= g_{\mathrm{HQ}} g_{lq/l \bar{q}}\left[\frac{64 \pi^2 \alpha^2}{9} \frac{\left(\left(s-m_{\mathrm{HQ}}^2\right)^2+\left(m_{\mathrm{HQ}}^2-u\right)^2 + 2 m_{\mathrm{HQ}}^2 (\frac{t^{2}+\gamma_{\mathrm{G}}^4}{t} )\right)t^{2}}{\left(t^{2}+\gamma_{\mathrm{G}}^4\right)^2}\right] \, .
\end{aligned}
\end{equation}
\end{widetext}
Here, $g_{\mathrm{HQ}} = N_{s}\times N_{c},\,  g_{g} = N_{s} \times (N_{c}^{2}-1)$ and $g_{lq/l \bar{q}} = N_{s}\times N_{c} \times N_{f}$ are the degeneracy factor for $\mathrm{HQ}$, gluon, and light quark respectively with $N_{s}=2, N_{f} =3$ and $N_{c} = 3$ have been used. 

\subsection{Radiative Process:}
In general, the transport coefficient can be written~\cite{Mazumder:2013oaa} as
\begin{equation}\label{tran_coeff._def.}
\text{X(p)} = \int \text{Phase space} \times \text{interaction}\times \text{transport part}
\end{equation}
The Eq.~\eqref{tran_coeff._def.} can be used in order to study the radiative contribution of the drag and diffusion coefficient by replacing the two-body phase space and invariant amplitude with their three-body counterparts, keeping the transport part the same~\cite{Mazumder:2013oaa}. Let us consider the $2\rightarrow 3 $ inelastic process: $\mathrm{HQ}(P)+l(Q) \rightarrow$ $\mathrm{HQ}\left(P^{\prime}\right)+l\left(Q^{\prime}\right) + g(K^{\prime})$, where $K^{\prime} = (E_{k^{\prime}},\bm{k}_{\perp}^{\prime},k_{z}^{\prime})$ is the four-momentum of the emitted soft-gluon by $\mathrm{HQ}$ in the final state. The general expression for the thermal averaged $\llangle F(\bm{p})\rrangle $ for $2$ $\rightarrow$ $3$ process is given by~\cite{Mazumder:2013oaa}
\begin{figure}
\centering
\subfloat[\label{fig2(a)}]{\includegraphics[scale=1.2]{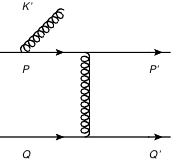}}
\quad \quad \quad
\subfloat[\label{fig2(b)}]{\includegraphics[scale=1.2]{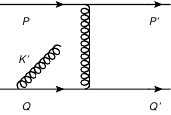}}
\quad \quad \quad
\subfloat[\label{fig2(c)}]{\includegraphics[scale=1.2]{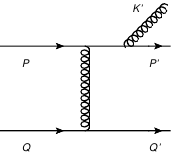}}
\quad \quad \quad
\subfloat[\label{fig2(d)}]{\includegraphics[scale=1.2]{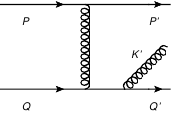}}
\quad \quad \quad
\subfloat[\label{fig2(e)}]{\includegraphics[scale=1.2]{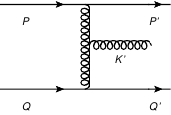}}
\captionof{figure}{Feynman diagrams for process $\mathrm{HQ}(P)+l(Q) \rightarrow$ $\mathrm{HQ}\left(P^{\prime}\right)+l\left(Q^{\prime}\right) + g(K^{\prime})$, showing an inelastic scattering of $\mathrm{HQ}$ with light quark and a soft gluon emission.}
\label{fig_2}
\end{figure}	
\ba
\llangle F(\bm{p})\rrangle_{\mathrm{rad}} & \hspace{-0.1cm}=& \frac{1}{2 E_p g_{\mathrm{HQ}}}\int\hspace{-0.1cm} \frac{d^3 \bm{q}}{(2 \pi)^3 E_q} \int\hspace{-0.1cm}\frac{d^3 \bm{q}^{\prime}}{(2 \pi)^3 E_{q^{\prime}}}\int\hspace{-0.1cm} \frac{d^3 \bm{p}^{\prime}}{(2 \pi)^3 E_{p^{\prime}}} \nn
&\hspace{-0.4cm}\times & \hspace{-0.3cm} \int\hspace{-0.1cm}\frac{d^3 \bm{k}^{\prime}}{(2 \pi)^3 E_{k^{\prime}}} \sum\left|\mathfrak{M}_{2 \rightarrow 3}\right|^2  f_k\left(E_q\right)\left(1 \pm f_k\left(E_{q^{\prime}}\right)\right) \nn
&\hspace{-0.4cm}\times &  \left(1+f_g\left(E_{k^{\prime}}\right)\right) \theta_1\left(E_p-E_{k^{\prime}}\right) \theta_2\left(\tau-\tau_F\right) \nn
&\hspace{-0.4cm}\times & F(\bm{p})(2 \pi)^4\delta^{(4)} \left(P+Q-P^{\prime}-Q^{\prime}-K^{\prime}\right) \, .
\label{rad._process}
\ea
where $\tau$ is the scattering time of $\mathrm{HQ}$ with the medium constitutents and $\tau_{F}$ is the formation time of gluons. The theta function $\theta_{1}(E_{p}-E_{k^\prime})$ in Eq.~\eqref{rad._process} imposes the constraints on the process that the emitted gluon energy should be less than the initial energy of $\mathrm{HQ}$. Whereas, the second theta function $\theta_{2}(\tau-\tau_{F})$ makes sure that the formation time of gluon should be lesser than the scattering time of $\mathrm{HQs}$ with medium constituents that accounts for the Landau-Pomerancguk-Migdal (LPM) effect~\cite{Wang:1994fx, Gyulassy:1993hr, Klein:1998du}. Also, $f_{g}(E_{k^\prime}) = 1/[\exp{(\beta E_{k^\prime})}-1]$ is the distribution of the emitted gluon where $\beta = 1/T$ i.e. the gluons in Fig.~\ref{fig:1} and Fig.~\ref{fig_2} are in thermally equilibrated state. The term $\left|\mathfrak{M}_{2 \rightarrow 3}\right|^2$ denotes the matrix element squared for the $2 \rightarrow 3$ radiative process as depicted in Fig.~\ref{fig_2}. It can be expressed in terms of the matrix element of the collision process multiplied by the probability for soft gluon emission~\cite{Abir:2011jb} as follows
\begin{equation}
\left|\mathfrak{M}_{2 \rightarrow 3}\right|^2=\left|\mathfrak{M}_{2 \rightarrow 2}\right|^2 \times \frac{12 g_s^2}{{k}_{\perp}^{\prime 2}}\left(1+\frac{m_{\mathrm{HQ}}^2}{s} e^{2\eta}\right)^{-2} \, ,
\end{equation}
where $g_{s}$ is the strong running coupling defined at one-loop as 
\begin{equation}
g_{s}^{2} (T)= 4\pi \alpha_{s} = \frac{24 \pi^{2}}{11N_{c}-2N_{f}}\frac{1}{\ln (2\pi T/\Lambda_{\overline{\text{MS}}})} 
\end{equation}
having scale $\Lambda_{\overline{\text{MS}}}=0.176$ GeV~\cite{Haque:2014rua} for $N_{f}=3$, $\eta$ is the rapidity of the emitted gluon and $\left(1+\frac{m_{\mathrm{HQ}}^2}{s} e^{2 \eta}\right)^{-2}$ is the suppression factor for the $\mathrm{HQ}$ due to the dead-cone factor~\cite{Dokshitzer:2001zm, Abir:2011jb}. From the Eq.~\eqref{rad._process} we have
\begin{equation}
\begin{aligned}
\llangle F(\bm{p})\rrangle_{\mathrm{rad}}= & \llangle F(\bm{p})\rrangle_{\mathrm{coll.}} \times \mathfrak{I}(\bm{p}) \, ,
\end{aligned}
\end{equation}
where $\mathfrak{I}(\bm{p})$ is given by
\ba
\mathfrak{I}\left(\bm{p}\right) &= & \int \frac{d^3 k^{\prime}}{(2 \pi)^3 2 E_{k^{\prime}}} \frac{12 g_s^2}{k_{\perp}^{\prime 2}}\left(1+\frac{m_{\mathrm{HQ}}^2}{s} e^{2 \eta}\right)^{-2} \nn
&\times& \left(1+f_g\left(E_{k^{\prime}}\right)\right) \theta_1\left(E_p-E_{k^{\prime}}\right) \theta_2\left(\tau-\tau_F\right) \, .
\label{Int_I_k}
\ea
In the limit of soft gluon emission ($\theta_{k^{\prime}} \rightarrow 0$), one will get $ \left(1+\frac{m_{\mathrm{HQ}}^2}{s} e^{2 \eta}\right)^{-2} \approx\left(1+\frac{4 m_{\mathrm{HQ}}^2}{s \theta_{k^{\prime}}^2}\right)^{-2},$ where $\theta_{k^{\prime}}$ is the angle between the radiated soft gluon and the $\mathrm{HQ}$ which can be related to the rapidity parameter through the relation $\eta=-\ln \left[\tan \left(\theta_{k^{\prime}} / 2\right)\right]$. 
In order to simplify the equation~\eqref{Int_I_k}, one can convert emitted gluon four-momentum in terms of the rapidity variable as
\begin{equation}
\begin{aligned}
E_{k^{\prime}} = k_{\perp}^{\prime} \cosh\eta \, , \quad \quad k_{z}^{\prime} = k_{\perp}^{\prime} \sinh\eta \, ,
\end{aligned}
\end{equation}
with $d^{3}k^{\prime} = d^{2}k_{\perp}^{\prime}dk_{z}^{\prime} = 2\pi k_{\perp}^{\prime 2}dk_{\perp}^{\prime}\cosh\eta \,d\eta$. The interaction time $\tau$ is related to the interaction rate $\Gamma = 2.26 \alpha_{s}T$~\cite{Shaikh:2021lka} and the $\theta_{2}(\tau-\tau_{F})$ impose the constraint
\begin{equation}
\begin{aligned}
\tau = \Gamma^{-1} > \tau_{F} = \frac{\cosh\eta}{k_{\perp}^{\prime}} \, ,
\end{aligned}
\end{equation}
which shows that $k_{\perp}^{\prime} > \Gamma \cosh\eta = (k_{\perp}^{\prime})_{\text{min.}}$. Further, from the other theta function $\theta_{1}(E_{p}-E_{k^{\prime}})$ we have, 
\begin{equation}
\begin{aligned}
E_{p} > E_{k^{\prime}} = k_{\perp}^{\prime} \cosh\eta \, , \quad \quad (k_{\perp}^{\prime})_{\text{max.}} = \frac{E_{p}}{\cosh\eta} \, .
\end{aligned}
\end{equation}
Also, the Bose enhancement factor for the emitted gluon in the limiting case ($E_{k^{\prime}} \ll T$) can be written as
\begin{equation}
\begin{aligned}
1+f_g\left(E_{k^{\prime}}\right)
=\frac{T}{k_{\perp}^{\prime} \cosh\eta} \, .
\end{aligned}
\end{equation}
Thus the integral $\mathfrak{I}(\bm{p})$ becomes
\ba
\mathfrak{I}\left(\bm{p}\right) &= & \, \frac{3}{2 \pi^2} g_s^2 T \int_{\Gamma \cosh \eta}^{E_p / \cosh \eta} d k_{\perp}^{\prime} \int_{-\eta_{1}}^{\eta_{1}} d \eta \nonumber \\
& \times & \left(1+\frac{m_{\mathrm{HQ}}^2}{s} e^{2 \eta}\right)^{-2} \frac{1}{k_{\perp}^{\prime} \cosh \eta} \, ,
\ea
where rapidity integration limits are decided based on the pseudo-rapidity coverage of the detector accordingly. In the next section we have used the value of $\eta_{1}=20$ for practical calculations.
\section{Results and Discussion}\label{sec.3}

\begin{figure}
\centering
{\includegraphics[scale=0.65]{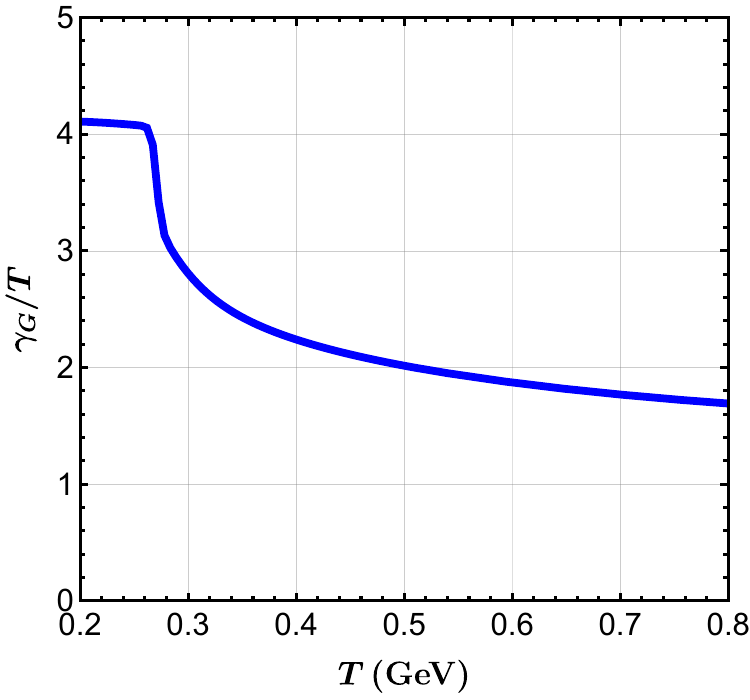} }
\caption{Temperature dependence of the scaled Gribov mass parameter obtained by matching the thermodynamics of the quasi-particle approach with the pure gauge lattice  data~\cite{Borsanyi:2012ve}.}
\label{gammaG_variation}
\end{figure}

In order to do a numerical evaluation of drag and diffusion coefficients, firstly, we must fix the Gribov mass parameter $\gamma_{\mathrm{G}}$ appearing in the Gribov propagator. To do so, the authors of~\cite{Jaiswal:2020qmj} have done the matching of temperature-dependent scaled trace anomaly results of lattice~\cite{Borsanyi:2012ve} with the equilibrium thermodynamic quantities. In Fig.~\ref{gammaG_variation}, we showed the scaled Gribov mass parameter variation $\gamma_{\mathrm{G}}/T$ with temperature $T$. This dependence of $\gamma_{\mathrm{G}}$ will be used in the estimation of other quantities evaluated further.
Now, we will present our numerical results for the transport coefficient, namely drag and diffusion for elastic and inelastic processes, the specific shear viscosity of the QGP medium, and the estimation of collisional and radiative energy loss in the separate subsections.

\subsection{Drag and diffusion coefficient for collisional and radiative processes}

\begin{figure}
\centering
\subfloat[\label{drag_temp.}]
{\includegraphics[scale=0.65]{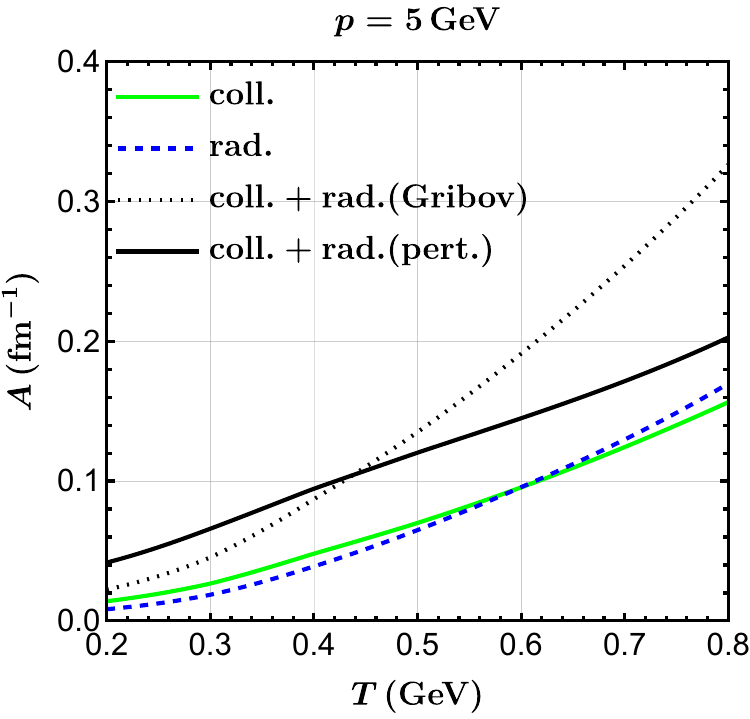} }
\quad \quad
\subfloat[\label{drag_mom.}]
{\includegraphics[scale=0.65]{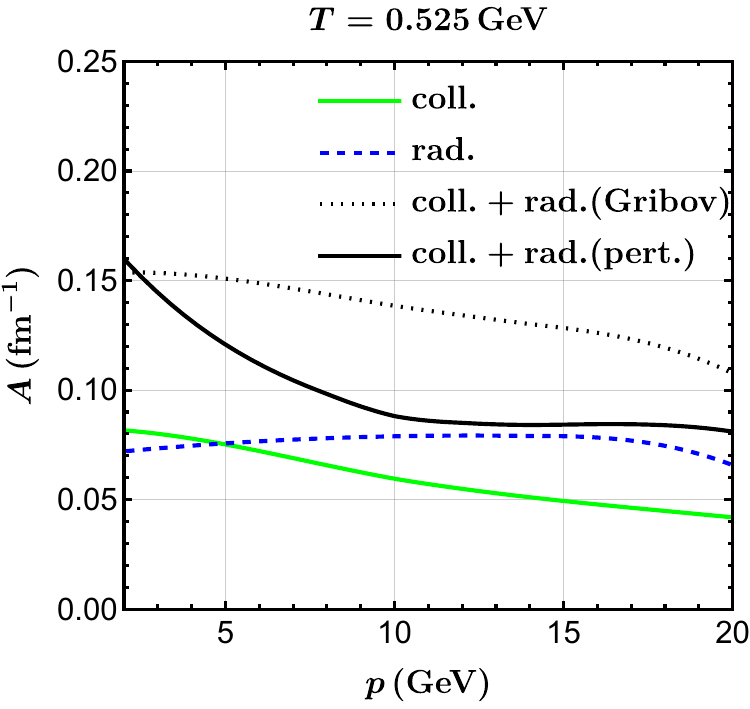} }
\vspace{.1cm}
\captionof{figure}{Variation of charm quark drag coefficient with temperature and momentum at (a) $p = 5$ GeV and (b) $T = 0.525$ GeV, respectively, where coll. stands for collisional
processes and rad. stands for radiative processes. The same abbreviations have been used for the rest of the Figs.}
\label{drag_coeff.}
\end{figure}

In Fig.~\ref{drag_temp.}, the temperature dependence of the drag coefficient has been shown at $p = 5$ GeV. Here, we have taken the charm quark mass $1.3$ GeV. The contributions from both processes have been shown as these processes occur independently in the thermal medium. Fig.~\ref{drag_temp.} shows that the collisional process contributes more at the low temperature than the radiative one. However, as the temperature increases, the radiative process starts dominating, indicating that inelastic processes are more important at LHC energies than RHIC energy within this model calculations. As the temperature increases, the total contribution to the drag coefficient increases compared to the elastic process. Qualitatively, the drag coefficient has a similar nature within this modeling compared to earlier perturbative results~\cite{Mazumder:2013oaa}. However, the overall magnitude of both processes is higher after a $T=0.4$ GeV and lower before $T=0.4$ GeV, which can be inferred from the non-perturbative nature of the Gribov propagator. As the system with the Gribov gluon propagator is strongly interacting, the $\mathrm{HQs}$ feel a strong drag force compared to the weakly interacting matter in the high-temperature domain while at lower temperature $\mathrm{HQ}$ drag coefficient is less compared to earlier perturbative estimation. In other words, one would expect a larger drag coefficient in Gribov plasma for large temperatures and a lower drag coefficient for lower temperatures. Thus, the overall magnitude of the drag coefficient for collisional and radiative processes is higher in the high-temperature domain via the Gribov-Zwanziger approach than it was with earlier perturbative results.
In Fig.~\ref{drag_mom.}, the drag coefficient of $\mathrm{HQ}$ has been plotted with its momentum for a temperature $T = 0.525$ GeV. It has been observed that after a momentum of $5$ GeV, the radiative contribution dominates in the medium despite the dead cone effect. 

\begin{figure}
\centering
\subfloat[\label{tran._diff._temp.}]
{\includegraphics[scale=0.65]{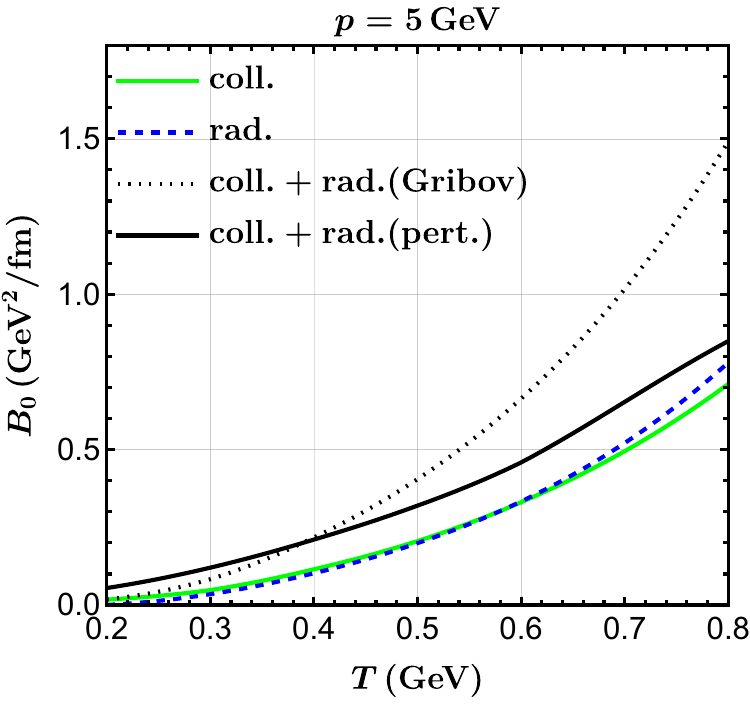} }
\quad \quad
\subfloat[\label{tran._diff._mom.}]
{\includegraphics[scale=0.65]{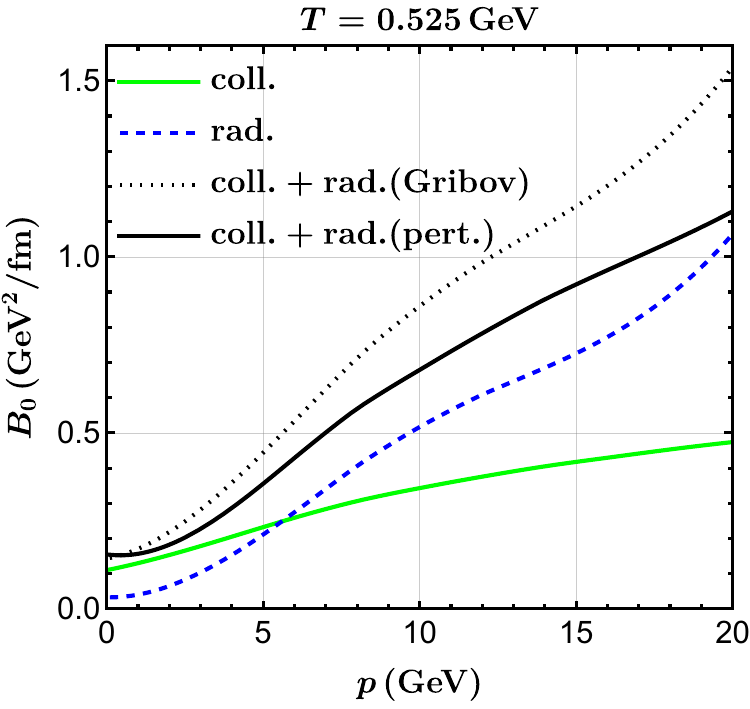} }
\vspace{.1cm}
\captionof{figure}{Variation of charm quark transverse diffusion coefficient with temperature and momentum at (a) $p = 5$ GeV and (b) $T = 0.525$ GeV respectively.}
\label{diff._tran.}
\end{figure}

\begin{figure}
\centering
\subfloat[\label{long._diff._temp.}]
{\includegraphics[scale=0.65]{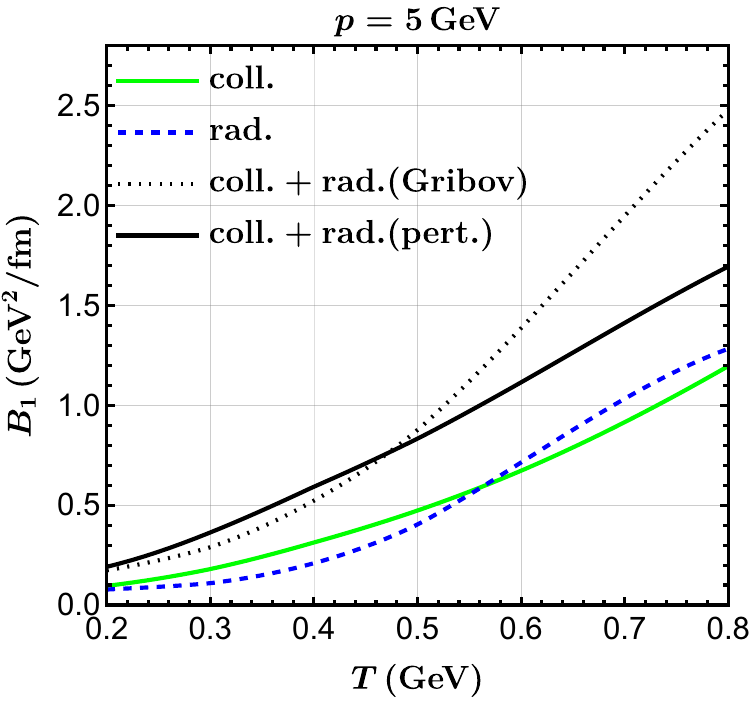} }
\quad \quad
\subfloat[\label{long._diff._mom.}]
{\includegraphics[scale=0.65]{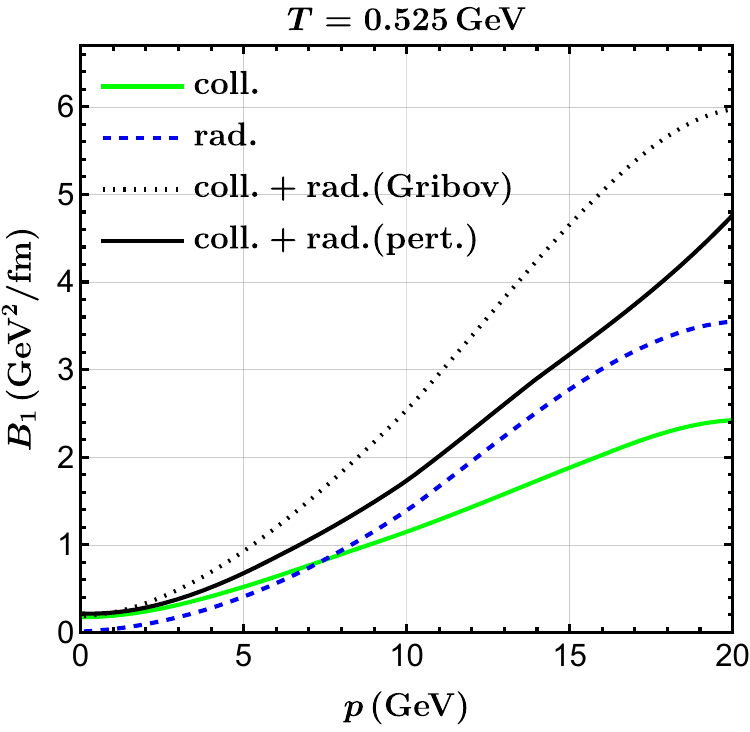} }
\vspace{.1cm}
\captionof{figure}{Variation of charm quark longitudinal diffusion coefficient with temperature and momentum at (a) $p = 5$ GeV and (b) $T = 0.525$ GeV respectively.}
\label{diff._long.}
\end{figure}

In Figs.~\ref{tran._diff._temp.} and~\ref{long._diff._temp.}, the variation of the transverse and longitudinal diffusion coefficient of the charm quark is shown with respect to temperature. Similar to the drag coefficient, the radiative effects start dominating for the high-temperature range around $T = 0.6$ GeV. The transverse and the longitudinal diffusion coefficient have a smaller magnitude before $T\sim 0.45$ GeV and a larger magnitude after $T\sim 0.45$ GeV compared to earlier perturbative results~\cite{Mazumder:2013oaa}, pertaining to the more non-perturbative nature. Similarly, Fig.~\ref{tran._diff._mom.} and \ref{long._diff._mom.} shows the transverse and longitudinal diffusion coefficient variation with the charm momentum at $T = 0.525$ MeV. The variation with the momentum $p$ for the transverse diffusion is smaller than the longitudinal diffusion coefficient. Although the nature of the diffusion coefficient variation differs from the drag coefficient, the radiative effects dominations after $p = 5$ GeV are clearly evident, showing the importance of radiative effects at high momenta.    
\\
A similar analysis of drag and diffusion coefficients can be done for bottom quarks having mass approximately as $4.2$ GeV easily. We report that for the bottom quarks, the drag coefficient magnitudes decrease compared to the charm quark drag coefficient magnitude because of the greater mass of the bottom quark compared to the charm quark. Similar behavior is also observed for the transverse and longitudinal diffusion coefficients as well.
\subsection{Shear viscosity to entropy density ratio ($\eta/s$) of QGP}
\begin{figure}
\centering
{\includegraphics[scale=0.7]{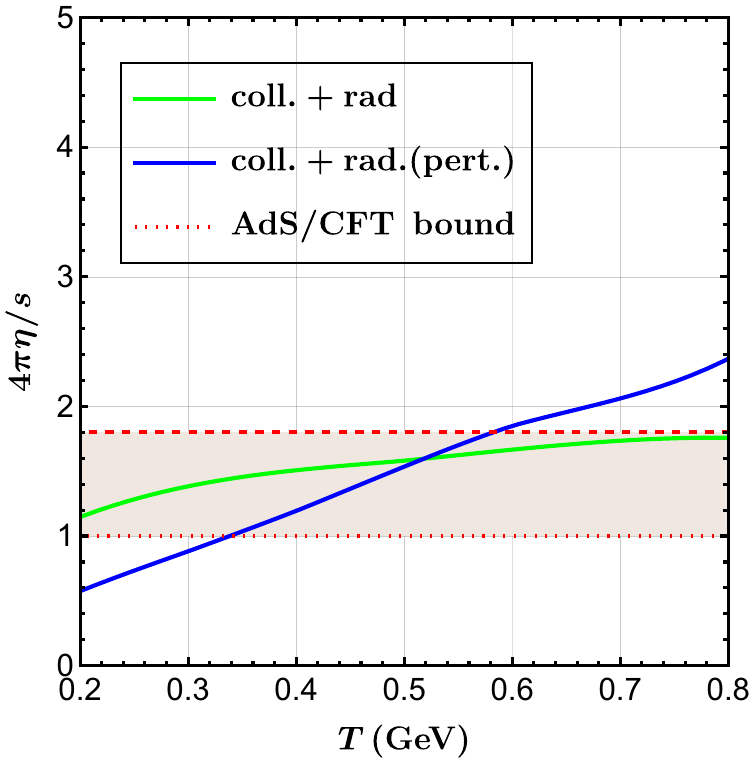} }
\vspace{.1cm}
\captionof{figure}{The value of $4\pi \,\eta /s$ for a charm quark with momentum $\langle p_{z}\rangle$ $= 5$ GeV propagating in QGP medium of temperature $T$.}
\label{shear_viscosity}
\end{figure}

In this subsection, we estimate the shear viscosity to entropy density ratio by using the Gribov propagator, which enters the interaction part of the diffusion coefficient. The transverse momentum diffusion coefficient $B_{0}$ can be written as
\begin{equation}
B_{0}=\frac{1}{2}\left(\delta_{i j}-\frac{p_i p_j}{p^2}\right) B_{i j}  \, ,
\end{equation}
By using Eq.~\eqref{diffusion_definition} and putting $(p^{\prime}-p)_{i} = k_{i}$,
\begin{equation}
\begin{aligned}
B_{0} = \frac{1}{4} \llangle[\Bigg]\left({k}^{2} - \frac{(\bm{p}\cdot\bm{k})^{2}}{{p}^{2}}\right)\rrangle[\Bigg]\, ,
\end{aligned}
\end{equation}
If $\mathrm{HQ}$ momentum is considered in the $\hat{z}$ direction, then 
\begin{equation}
\begin{aligned}
B_{0} = \frac{1}{4} \llangle[\big]k_{\perp}^{2} \rrangle[\big]\ = \frac{1}{4} \hat{q} \, .
\end{aligned}
\end{equation}
where $\hat{q}$ is the jet quenching parameter, which is also an important quantity for the characterization of QGP. Recently, the relation between these two parameters, namely specific shear viscosity $\eta/s$ and dimensionless quenching parameter $\hat{q}/T^{3}$ has been calculated up to next-to-leading order in terms of coupling constant using perturbative QCD approach in~\cite{Muller:2021wri}. Thus, we estimated $\eta/s$ of QGP using the following expression
\begin{equation}
\frac{\eta}{s} = 1.63 \frac{T^3}{\hat{q}} \, ,
\end{equation}
Thus,
\begin{equation}
4 \pi \frac{\eta}{s} = 1.63 \pi \frac{T^3}{B_{0}} \, .
\end{equation}
In Fig.~\ref{shear_viscosity}, we plotted $4\pi \,\eta/s$ with respect to temperature $T$ within this model calculations. We compared it with the standard KSS bound having values of $4\pi \eta/s = 1.0 - 1.8$ as obtained in~\cite{Kovtun:2004de}, as well with the earlier perturbative result obtained in~\cite{Mazumder:2013oaa}. The obtained results show that the value of $4\pi \, \eta/s$ comes strictly within the AdS/CFT bound after the inclusion of radiative processes, which further improves the earlier perturbative results and shows a good agreement with the experimental values~\cite{Bernhard:2019bmu, JETSCAPE:2020mzn}. For the earlier perturbative results Debye mass $(m_{D})$ acts as a infrared regulator and the value of $m_{D}=\sqrt{3/2}g_{s}T$ is used in Figs.~\ref{drag_coeff.},~\ref{diff._tran.}
,~\ref{diff._long.} and \ref{shear_viscosity} for the perurbative result comparison with GZ estimations. Thus, one can infer that the Gribov-Zwanziger technique improves the perturbative results in the low-temperature domain as well as in the high-temperature domain as can be observed in Fig.~\ref{shear_viscosity}.  
%

\subsection{Collisional and radiative energy loss:}
\begin{figure}
\centering
\subfloat[\label{Energy_loss_360MeV}]
{\includegraphics[scale=0.65]{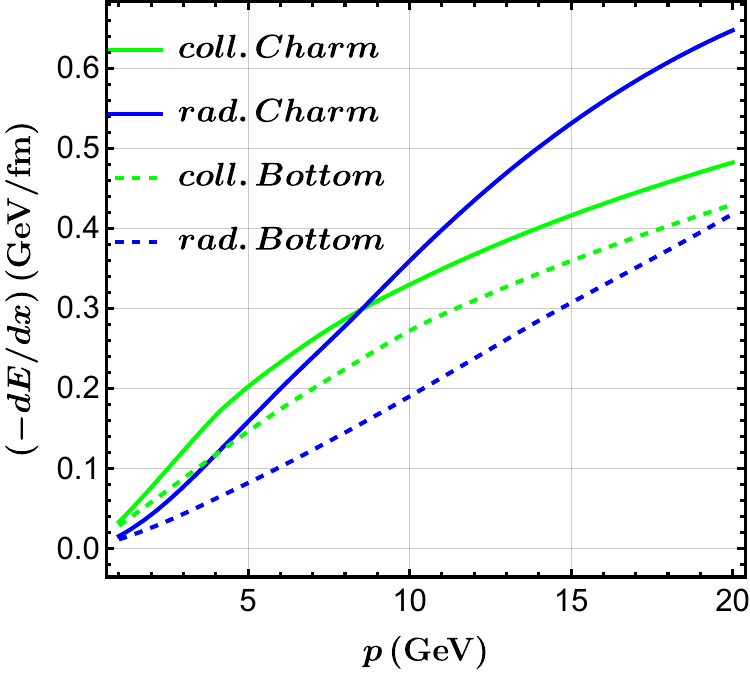} }
\quad \quad
\subfloat[\label{Energy_loss_480MeV}]
{\includegraphics[scale=0.65]{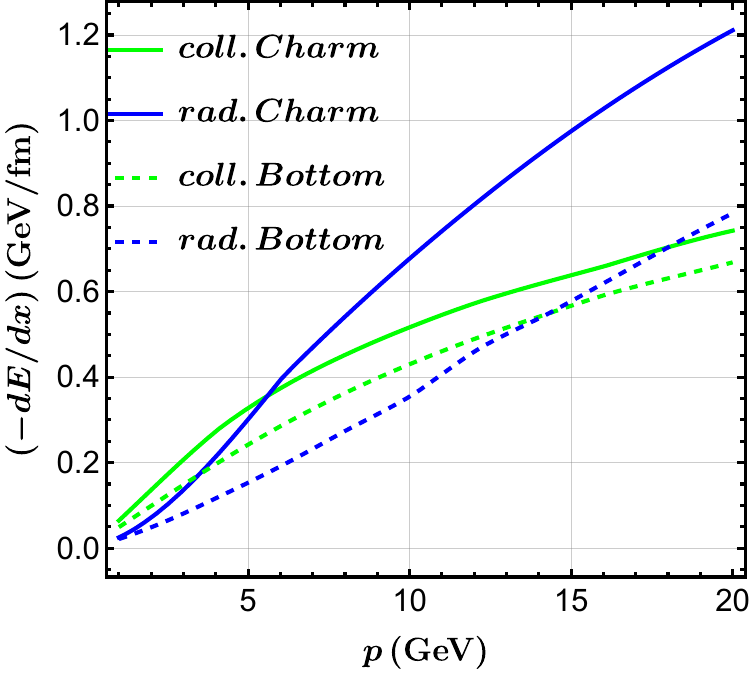} }
\vspace{.1cm}
\captionof{figure}{ Momentum variation of elastic and radiative energy loss of the $\mathrm{HQs}$ in the medium for the RHIC energy at $T = 360 $ MeV (upper panel) and the LHC energy at $T = 480 $ MeV (lower panel). Solid lines are for charm quarks, while dotted ones are for bottom quarks.}
\label{energy_loss}
\end{figure}

The differential energy loss of the $\mathrm{HQ}$ is related to the drag coefficient~\cite{GolamMustafa:1997id} and can be expressed as
\begin{equation}
-\frac{dE}{dx} = A(p^{2},T) \,p \,. 
\end{equation}
In Fig.~\ref{energy_loss}, we have plotted the energy loss of $\mathrm{HQs}$ with respect to their momentum $p$, showing collisional and radiative contributions independently at RHIC and LHC energies. In Fig.~\ref{Energy_loss_360MeV}, the energy loss at RHIC energy ($T = 0.36$ GeV) for charm (solid lines) and bottom (dotted lines) quarks are shown. Similarly, Fig.~\ref{Energy_loss_480MeV} has been plotted for temperature $T = 0.48$ GeV, i.e., at LHC energy. As expected, energy loss for the bottom quark is less compared to the charm quark because of more drag offered to the bottom quark in the medium due to its large mass. Due to restricted phase space, the collisional processes dominate in the initial momentum range around $5$ GeV. However, after that, the radiative process dominates the collisional one for charm quark at both energies at LHC and RHIC. In the case of the bottom quark, the collisional process contribution dominates in the whole momentum range at RHIC. At the same time, this nature continues at LHC energy until $\sim 15$ GeV, then the radiative process dominates. This suppression in the radiative energy loss in the case of the bottom quark in comparison to the charm quark can be accounted for because of the dead cone factor, which prohibits the $\mathrm{HQ}$ from radiating gluon at a small angle. Thus, the dead-cone angle will be large for the higher mass, and the probability of energy loss due to radiation will be lesser.  

\section{Summary and Conclusion}\label{sec.4}
In this work, we have investigated the momentum and temperature dependence of the drag and diffusion coefficient of $\mathrm{HQs}$ propagating in the QGP medium. These transport coefficients play a pivotal role in the $\mathrm{HQ}$ phenomenology as they essentially govern the dynamics of the $\mathrm{HQs}$ in the Fokker-Planck approach. In the present work, our primary focus has been to incorporate the non-perturbative effects in the estimation of drag and diffusion coefficients, especially in the temperature regime close to the crossover. For this purpose, we take recourse to the Gribov-Zwanziger method. In this framework, the gluon propagators present in the scattering amplitudes have been replaced by the Gribov-modified propagators. It should be noted here that it has been a standard practice to include the Debye mass ($m_{D}\sim g(T)T$) as an infrared regulator in the $t-$channel matrix amplitude in order to circumvent the infrared divergence. However, in the present work, the mass scale in the modified gluon propagator arises naturally within the model framework, resulting in a finite $t-$channel contribution. The temperature dependence of the mass scale has been extracted by matching the thermodynamics of the Gribov plasma with the pure gauge lattice results. Once the temperature dependence has been fixed, we incorporate this modified gluon propagator in the collisional and radiative contributions to obtain the momentum and temperature dependences of the drag and diffusion coefficient, which show a significant increment compared to earlier perturbative estimations. 
Moreover, we find that the estimation of the specific shear viscosity using the Gribov method is in better agreement with the AdS/CFT calculations. Finally, we have investigated the collisional and radiative energy loss of charm and bottom quark traversing through the medium. We find that the energy loss in both cases is higher in magnitude compared to the earlier perturbative estimations. 

It should be mentioned here that the Gribov framework presented in this work is a simplistic approach to incorporate the non-perturbative effects relevant near the phenomenologically accessible temperature regime. Nevertheless, the present study serves as an important first step towards estimating the impact of Gribov-like approaches on the $\mathrm{HQ}$ dynamics and encourages one to study further different experimental observables like nuclear modification factor $R_{AA}$~\cite{Debnath:2023zet}, elliptic flow $v_{2}$ as well as other transport properties of the medium~\cite{Florkowski:2015dmm, Florkowski:2015rua, Jaiswal:2020qmj,Madni:2024xyj,Madni:2024ubw} within this framework. An interesting future direction in this regard would be to incorporate the dissipative effects in the estimation of drag and diffusion tensor of the HQ ~\cite{Srivastava:2016igg, Prakash:2021lwt, Kumar:2021goi, Prakash:2023wbs, Singh:2023smw}. Also, the non-trivial backgrounds like the strong external magnetic field may have a significant impact ~\cite{Fukushima:2015wck, Bandyopadhyay:2021zlm, Bandyopadhyay:2023hiv} on the transport properties of the Gribov modified plasma medium. We relegate such studies for future explorations. 

\begin{acknowledgements}
We would like to thank Raktim Abir for the helpful discussion. N.H. is supported in part by the SERB-Mathematical Research Impact Centric Support (MATRICS) under Grant No. MTR/2021/000939.
\end{acknowledgements}

\end{document}